\documentclass[fleqn,twoside]{article}
\usepackage{espcrc2}

\usepackage{graphicx}
\usepackage[figuresright]{rotating}

\newcommand{\rf}[1]{(\ref{#1})}
\newcommand{\beq}{\begin{equation}}
\newcommand{\eeq}{\end{equation}}
\newcommand{\bea}{\begin{eqnarray}}
\newcommand{\eea}{\end{eqnarray}}

\newcommand{\e}{\mbox{e}}
\renewcommand{\d}{\mbox{d}}
\newcommand{\g}{\gamma}
\newcommand{\G}{{\Gamma}}
\renewcommand{\l}{\lambda}
\renewcommand{\L}{\Lambda}
\renewcommand{\b}{\beta}
\renewcommand{\a}{\alpha}
\newcommand{\n}{\nu}
\newcommand{\m}{\mu}

\renewcommand{\th}{\theta}
\newcommand{\del}{\delta}
\newcommand{\Del}{\Delta}

\newcommand{\oh}{\frac{1}{2}}
\newcommand{\oq}{\frac{1}{4}}
\newcommand{\dg}{\dagger}

\newcommand{\tr}{\mbox{Tr}\;}
\newcommand{\ra}{\rangle}
\newcommand{\la}{\langle}

\newcommand{\mi}{\!-\!}
\newcommand{\equ}{\!=\!}
\newcommand{\pl}{\!+\!}

\newcommand{\cU}{{\cal U}}

\newcommand{\bpsi}{\bar{\psi}}

\newcommand{\AmS}{{\protect\the\textfont2
  A\kern-.1667em\lower.5ex\hbox{M}\kern-.125emS}}

\title{Strings, quantum gravity and non-commutative geometry on the lattice}

\author{Jan Ambj\o rn\address[MCSD]{Niels Bohr Institute, \\ 
        Blegdamsvej 17, DK-2100 Copenhagen \O, Denmark}%
        \thanks{This work was supported by MaPhySto,  Center for 
Mathematical Physics and Stochastics financed by the 
Danish National Research Foundation,  by 
EU network on ``Discrete Random Geometry'', grant HPRN-CT-1999-00161, 
and by ESF network no.82 on ``Geometry and Disorder''.}
}
       
\begin{document}

\begin{abstract}
I review recent progress in understanding non-perturbative aspects 
of string theory, quantum gravity and non-commutative geometry 
using lattice methods.
\vspace{1pc}
\end{abstract}

\maketitle

\section{Introduction}

Lattice methods have been  very successful dealing with string theory,
quantum gravity and non-commutative field theory 
in the following sense: whenever the bosonic 
theory is known to exist, the lattice formulation has \\
(1) provided a rigorous definition of the theory,\\
(2) a scaling limit where the continuum theory 
can be defined and which can be analyzed by numerical and/or 
analytical methods,\\
(3) often been superior to a continuum approach even 
from an analytical point of view.

Examples are 2d quantum gravity and non-critical strings with
a central charge $c \leq 1$ , non-commutative Yang-Mills theory 
(which was in fact discovered on the lattice as early as  1983
\cite{go,gk}), and 3d quantum gravity formulated as a Turaev-Viro 
state-sum. There are also examples where lattice attempts to 
formulate a non-perturbative theory have failed. But with
the advantage of hindsight we now understand that the 
message from the lattice was correct even in these cases. 
One such example is the
bosonic string theory in dimensions $d >1$ \cite{kkm,david,adf,adfo}. 
Probably there 
exists no non-tachyonic bosonic string theory in dimensions $d >1$.

The examples mentioned are non-trivial in the sense that they 
deal with diffeomorphism-invariant theories and one  
would not expect the lattice to be the best way to regularize 
such theories. In fact, the situation seems to be the opposite: the 
lattice framework introduced by the use of  ``dynamical triangulations''
seems to work very well for the above-mentioned bosonic theories.

However, there are two fundamental areas where the lattice 
approach cannot yet claim success: four-dimensional quantum
gravity and superstring theory. In the first case we 
do not know if there exists a consistent non-perturbative
theory of quantum gravity, formulated entirely in terms of 
the gravitational fields, or if the theory has to be embedded 
in a larger theory like superstring theory. However, {\it if} 
a purely bosonic theory of four-dimensional quantum gravity 
exists, it follows from the remarks above that the lattice 
framework should be ideal for a non-perturbative definition.
I will discuss a new lattice approach, 
called {\it Lorentzian simplical quantum gravity} in 
Sec.\ \ref{gravity}.
In the case of superstring theories,
lattice formulations have been stalled by the inability to 
implement supersymmetry in the correct way on the lattice.
While a number of attempts to implement worldsheet supersymmetry
on random lattices have failed, it has also been known that 
the chances  were better if one used a Green-Schwarz formalism.
In that case the supersymmetry is a {\it space-time} supersymmetry
rather than a world sheet supersymmetry and  thus not
in direct conflict with an underlying worldsheet 
lattice. The obstacle came in that case from the so-called
$\kappa$-symmetry which has not yet  been implemented as an exact
symmetry in the 
lattice approach, and it is unclear if it will automatically
be satisfied when the lattice spacing is taken to zero.

The so-called ``second string revolution'' which pointed 
to unexpected and sometimes non-perturbative 
connections between the five different superstring 
theories and the so-called $M$-theory, led to the 
quest for a {\it non-perturbative} definition of  
these theories. 
The {\it new} matrix model approach to $M$-theory and 
type IIB superstrings initiated by Banks, Fischler, Shenker and Susskind
(the BFSS matrix model \cite{bfss}), 
and by Ishibashi, Kawai, Kitazawa and Tsuchiya 
(the IKKT matrix model, \cite{ikkt}),
respectively, were attempts to provide such a definition.
In these discretized models 
supersymmetry was implemented in a way which 
avoided the problem with $\kappa$-symmetry.  
The ``continuuum limits'' of the 
models correspond to $N$, the size of the matrices, going
to infinity. However, even for finite $N$ the concept of 
supersymmetry has a well defined meaning in these models.
In this respect they differ from previous attempts to 
introduce supersymmetry and are more promising.

\section{Superstrings on the lattice}\label{string}  

\subsection{Theory}   

In the so-called Schild gauge the action of a type IIB 
superstring can be written as  
the following expression (in dimensions $D\equ 4,6$ and 10):
\beq\label{2.2}
S \equ \int \!\!\d^2\xi \sqrt{g} 
\Big( \oq \{ X^\m,X^\n\}^2 \mi \frac{i}{2} \bpsi \G_\m 
\{X^\m,\psi\}\Big),
\eeq
where $\{X,Y\}$ denotes the Poisson bracket between the 
variables $X(\xi_1,\xi_2)$ and $Y(\xi_1,\xi_2)$ and $\G_\m$ are 
suitably defined $\g$-matrices. The idea behind 
the IKKT matrix model is to replace worldsheet variables
by $N\! \times \! N$ matrices:
\beq\label{2.3}
X^\m(\xi_1,\xi_2) \to X_{ij}^\m,~~~\psi^\a(\xi_1,\xi_2) \to \psi_{ij}^\a,
\eeq
In this way 
the worldsheet coordinates $(\xi_1,\xi_2)$ are mapped into the 
matrix indices $(i,j)$ and the worldsheet is replaced by the 
``matrix lattice'' $(i,j)$. The continuum limit should 
be obtained in the limit $N \!\to \!\infty$. As usual when 
going from a classical theory to a quantum theory we 
replace the Poisson brackets by commutators:
\beq\label{2.4}
\{X,Y\} \to -i [X,Y],
\eeq
and in this way the IKKT action for the type IIB superstring 
becomes:
\beq\label{2.5}
S \equ -\tr\Big( \oq [X^\m,X^\n]^2\pl \oh \bpsi \G_\m [X^\m,\psi]\Big).
\eeq
The corresponding partition function is
\beq\label{2.6}
Z_N = \int \d \psi \d \bpsi \d X_\m \; e^{-S[\bpsi,\psi,X_\m]}.
\eeq 
As described by IKKT, this model has even for finite $N$ an 
invariance which might be called supersymmetry. The (bosonic part
of the) classical 
equations of motion is
\beq\label{2.7}
[X_\m,[X_\m,X_\n]]= 0,
\eeq
and the simplest solution is 
\beq\label{2.8}
[X_\m^{cl},X_\n^{cl}] = 0,~~~~(\psi = 0).
\eeq
This means that the matrices $X_\m^{cl}$ can be simultaneously 
diagonalized and the {\it spray} of eigenvalues in $R^{10}$
viewed as the points defining classical space-time.

The valleys $[X_\m,X_\n]\equ 0$ make the existence
of the integral \rf{2.6} non-trivial. 
The convergence of  the integral for 
all $N$ in dimensions $D=4,6,10$ (these are the 
dimensions $D > 3$ where classical  supersymmetry of the 
Green-Schwarz superstring can be formulated) was established in 
\cite{john}.

In a non-perturbative  closed string theory
{\it the dimensionality of real space-time should be determined
dynamically}. We would like 
to see that the typical ``quantum'' matrices $X_\m$ dominating the 
matrix integral \rf{2.6} are approximately 
diagonalizable and that the (suitably defined){\it spray of eigenvalues} 
in the large-$N$ limit constitutes a four-dimensional manifold: our world. 

In order to make this more quantitative one can define the 
``space-time'' uncertainty $\Del$ as a measure of the ten matrices
not being simultaneously diagonalizable:
\beq\label{2.9}
\Del^2 \equ \frac{1}{N} \Big( \tr X_\m^2 \mi
\max_{U \in SU(N)} \sum_i (U X_\m U^\dg)^2_{ii}\Big).
\eeq
$\Del^2\equ 0$ if the $X_\m$'s are simultaneously diagonalizable.
Let us take the $U\equiv U_{max}$ which minimizes the RHS of \rf{2.9}. 
We define the ``space-time coordinates'' 
\beq\label{2.10}
(x_\m)_i \equiv \Big( U_{max}X_\m U_{max}^\dg\Big)_{ii}.
\eeq
With this definition we can now define the {\it extension 
of space-time} $R$:
\beq\label{2.11}
R^2 \equ \frac{1}{N} \la \tr X_\m^2\ra = \frac{
\left\la\sum_{i<j} (x_i^\m-x_j^\m)^2\right\ra}{N(N-1)/2},
\eeq
as well as the density of the spray of space-time points
\beq\label{2.12}
\rho(r) = \frac{\left\la \sum_{i<j} 
\del (r-\sqrt{(x_i^\m-x_j^\m)^2}~\right\ra}{N(N-1)/2}.
\eeq
When $N \equ 2$ we have an $SU(2)$ matrix model and 
it has been proven that \cite{staudacher1} 
\beq\label{2.13}
\rho(r) \sim r^{-2D+5},~~~~r~{\rm large}.
\eeq
There exists good arguments in favor of \rf{2.13} 
for all values of $N$ \cite{staudacher1,bielefeld,john}.

The simplest way to probe the effective dimension 
of the space-time dynamically generated from the 
distribution of $x_i^\m$'s is to look at the 
``moments of inertia'' for such a space-time:
\beq\label{2.14}
T^{\m\n}\equ \frac{2}{N(N\mi 1)} \sum_{i<j} 
\left\la(x_i^\m\mi x_j^\m)(x_i^\n\mi x_j^\n)\right\ra.
\eeq
This is a $10\!\times\! 10$ matrix and the principal moments of 
inertia for the distribution of $x_i^\m$ 
are the eigenvalues $\l_1 \geq \cdots \geq \l_{10}$.

Our four-dimensional world should  appear 
as a flat four-dimensional pancake distribution of $x_i^\m$ in the 
large $N$ limit, i.e.\  the four first eigenvalues 
of $T_{\m\n}$ should separate from the rest such that  one has 
a spontaneous breakdown of the ten-dimensional Lorentz 
invariance in the large-$N$ limit. Further,  
one would like the ratio $\Del/R \to 0$ in the large 
$N$ limit, such that it makes sense to talk about 
a classical background.

The power behavior \rf{2.13} of $\rho(r)$ is a 
source of ambiguity in defining 
$R$  and $T_{\m\n}$, since $ \la \tr X^{2n} \ra \equ \infty$
for sufficiently large $n$. This is in contrast to the 
situation where $\rho(r)$ falls off exponentially.

\subsection{Numerical results}

\begin{figure}[htb]
\vspace{-.2cm}
\includegraphics[width=5.5cm,angle=-90]{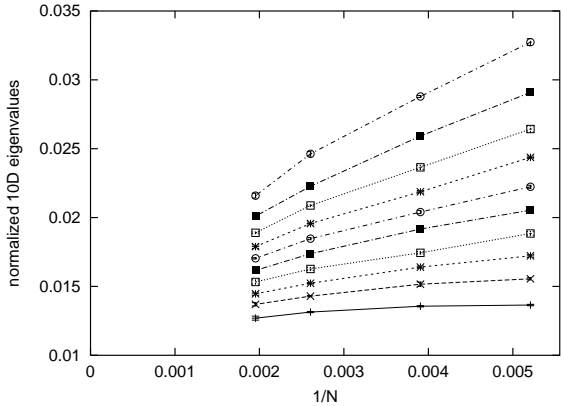}
\vspace{-1cm}
\caption[fig1]{The 10 eigenvalues of $T_{\m\n}$, extracted from 
Monte Carlo simulations of the one-loop approximation to 
\rf{2.15} with $M(X)$ replaced by $|M(X)|$, and plotted as a 
function of $1/N$.} 
\label{fig1}
\end{figure}

\begin{figure}[htb]
\includegraphics[width=7.5cm]{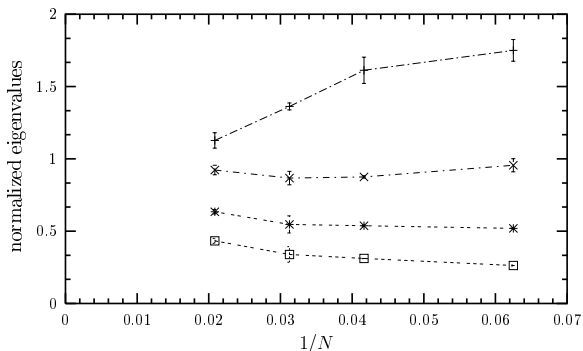}
\vspace{-1.5cm}
\caption[fig2]{The eigenvalues of $T_{\m\n}^{new}$
in the case $D\equ 4$, plotted as a function of $1/N$}
\label{fig2}
\end{figure}

Even if the non-perturbative regularization of the type IIB superstring
has resulted in a finite-dimensional ``lattice''-theory, the 
lattice being the entries of the $N\!\times\! N$ matrices $X_\m$,
the theory is not well suited for numerical simulations since 
it has fermions. One can integrate out the fermions from \rf{2.6}
and obtain
\beq\label{2.15}
Z_N = \int \prod_\m\d X_\m \; e^{ \oq \tr [X_\m,X_\n]^2}\; \det M(X),
\eeq
where $M(X)$ is a $(N^2\mi 1)\!\times\! (N^2\mi 1)$ matrix. For 
generic $X$ it is complex if the dimension $D$ of space-time is 
6 or 10, and real and positive if $D \equ 4$. The first question
we want to address is whether we observe any trace of spontaneous 
breakdown of Lorentz invariance. Analytical arguments in favor of  
a dimensional reduction to four dimensions have been given 
\cite{ikkt}, using a one-loop approximation,
but although very encouraging they need 
to be substantiated. Computer simulations, using  the same one-loop 
approximation to the action, can be performed if one drops the 
phase of the determinant in \rf{2.15}. Let us define
\beq\label{2.16}
\det M_\n(X) = |\det M(X)| \; e^{\n\G(X)}.
\eeq
$\n \equ 1$ corresponds to the physical situation we 
want to solve. $\n \equ 0$ corresponds to the situation we can 
simulate on the computer. There exist arguments in favor of 
a dimensional reduction  if $\n \equ \infty$ \cite{nv}.
Consequently, if we observe symmetry breaking
for $\n \equ 0$ it would be a strong argument in favor of symmetry breaking
for $\n\equ 1$. In Fig.\ \ref{fig1} we show the result of the 
measurement of the eigenvalues of $T_{\m\n}$ in the case of 
$\n \equ 0$. For details of the computer simulations shown 
here, see \cite{aabn}.
From the results shown in Fig.\ \ref{fig1} there is no trace of 
a spontaneous symmetry breakdown. This is an indication that 
the phase of the determinant may play a decisive role in 
a possible symmetry breakdown. However, in 
\cite{bielefeld} the model was 
investigated in $D \equ 4$ where the fermionic 
determinant is real and positive  and a dimensional reduction 
to {\it one dimension} was observed. This result highlights the 
ambiguity associated with power law distributions like 
\rf{2.13}. Indeed, $T_{\m\n}$ as defined in \rf{2.14}
is divergent if $D \equ 4$ and  one of  the eigenvalues 
of $T_{\m\n}$  diverges. According 
to \cite{bielefeld} this is a reflection of the 
fact that the {\it tail} of the distribution $\rho(r)$ is 
caused by aligned one-dimensional configurations of 
$x_i^\m$'s. It depends on  the choice of 
observables how sensitive they are to this tail. As an example
one can define a modified, converging $T_{\m\n}^{new}$ in 
$D\equ 4$:
\beq\label{2.17}
T_{\m\n}^{new} = \frac{2}{N(N\mi 1)} 
\sum_{i<j} \left\la \frac{(x_i^\m\mi x_j^\m)(x_i^\n\mi x_j^\n)}{
\sqrt{(x_i-x_j)^2}}\right\ra.
\eeq 
Using the definition \rf{2.17} one does not see any 
trace of spontaneously symmetry breaking in the 
case of $D \equ 4$. This is illustrated in Fig.\ \ref{fig2}. 
Also measurements of Wilson loops,
which have the interpretation as expectation values of 
fundamental strings in the theory, do not show any sign 
of spontaneous symmetry breaking \cite{aabn}. 

This leaves the question 
concerning the spontaneous breaking of  
Lorentz invariance in the model unsettled. More work is needed
in order to understand whether or not the signal 
of breaking depends on the observables chosen, and if it 
{\it does}, how to interpret this. In addition it would of course 
be very desirable if one could perform simulations in $D\equ 10$,
using the full complex determinant (or more precisely, Pfaffian).
It would require handling a complex action, a notorious problem
in computer simulations. Progress in dealing with a problem
of the kind encountered here has been been made recently \cite{an}.

\subsection{The BFSS matrix model}
The BFSS matrix model was conjectured to provide 
a non-perturbative definition of $M$-theory. Thus 
it is an eleven-dimensional theory where the 
matrices $X(t)$ depend on time. 
The problem with fermions also exists
in this model, and it is more difficult to simulate
than the IKKT model, since it is a matrix chain model,
not a single matrix model. A first attempt has been 
reported at this conference \cite{wosiek}. Finally, extensive 
numerical studies of a mean field
approximation to the BFSS model have been 
performed recently \cite{lifschytz}.

\section{Non-commutative gauge theories on the lattice}\label{non-commu}

Non-commutative Yang-Mills theory was discovered on the 
lattice \cite{go,gk}. In the context of the type IIB superstrings 
discussed in the last section they occur by expanding 
around a  classical solution to \rf{2.7} different from 
\rf{2.8} \cite{cds,nc-ikkt}, namely one satisfying
\beq\label{3.1}
[X_\m^{cl},X_\n^{cl}]= iC_{\m\n},
\eeq
where $C_{\m\n}$ are $c$-numbers. Note that \rf{3.1}
requires the matrices $X_\m^{cl}$ to be infinite-dimensional.
Thus one has to work directly at the $N \equ \infty$ limit.
Non-commutative field theories and in particular non-commutative
gauge theories have recently been studied intensively because 
they appear both in type IIB string theories and in 
open string theory (where the gauge theory, living on the 
$D$-branes where the open strings end, becomes non-commutative
in the presence of an external so-called $B_{\m\n}$ field).
But non-commutative gauge theories can be defined and studied 
independently of string theory and were indeed studied before 
they appeared in string theory.

As already noted when the theories were discovered,
a natural regularization of non-commutative YM theory  
is provided by the TEK lattice model \cite{go} by  
changing from matrices $X_\m$ to the exponentials 
$U_\m = \e^{ia X_\m}$. This has the following 
virtues: (1) It respects non-commutative gauge symmetry in the same 
way as ordinary gauge symmetry is respected on the lattice.
(2) It can be formulated for finite $N$ (this is what provides 
the regularization of the theory) and (3)
it preserves Morita equivalence.

The TEK model is  $U(N)$ gauge theory on a hypercube with 
twisted boundary conditions, given by the partition function
\bea
&& Z_{TEK}(U(N)) \equ  \int \!\! \prod_\m \d U_\m\nonumber \\
&&~~~~~\e^{ \frac{1}{2g^2} \sum_{\m <\n} \Big( 
Z_{\m\n}\tr U_\m U_\n U_\m^\dg U_\n^\dg \pl {\rm H.C.}\Big)}.
\label{3.2}
\eea
where $Z_{\n\m} \equ \e^{4\pi n_{\m\n}/N}\in Z_N$. 
The classical vacuum of the model 
becomes non-trivial and can be expressed in terms of the 
so-called twist eaters $\G_\m$: 
\beq\label{3.3}
U_\m^{cl}\equ \G_\m,~~~\G_\m\G_\n \equ Z_{\m\n}\G_\n\G_\m.
\eeq
This equation replaces \rf{3.1}.
Let $N \equ L^{D/2}$, $D$ even. Then the twist eaters can be 
used to construct the so-called Weyl map from the matrices
$U_\m^{ij}$  to functions $\cU_\m(x)$ with arguments $x$
on the hypercubic lattice $Z_L^D$. This is the inverse map 
to \rf{2.3}. By this map matrix multiplication is 
mapped to the star product of functions:
\beq\label{3.4}
U^{ij} \to \cU(x),~~\sum_k U_1^{ik}U_2^{kj} \to \cU_1(x) * \cU_2(x),
\eeq
where the star product $*$ is defined as 
\bea
A(x)*B(x)\equ \sum_{y,z} \e^{2i\th^{-1}_{\m\n} y_\m z_\n} 
A(x\pl y) B(x\pl z),
\eea
and $\th_{\m\n} \sim n_{\m\n}$ (see \cite{amns} for details).
The partition function
$Z_{TEK}(U(N))$ is mapped to 
\beq\label{3.5}
Z_{L^D}^{(NC)}(U(1)) = \int \prod_\m\d \cU_\m(x) \; \e^{-S^{(NC)}(\cU)},
\eeq
\bea
S^{(NC)}(\cU)\!\!\!\! & \equ&\!\!\! \frac{1}{2e^2} 
\!\sum_{\m<\n,x} \! \cU_\m(x)*\cU_\n(x\pl \hat{\m})* \nonumber\\
&&~~~~~~~~~~\cU_\m^\dg (x\pl \hat{\n}) * \cU_\n^\dg(x),
\eea
where $e^2 \equ g^2 N$. This relation:
\beq\label{3.7}
Z_{TEK}(U(N)) = Z_{L^D}^{(NC)}(U(1)),
\eeq
which states that a $U(N)$ commutative gauge theory on a 
hypercube (a $1^D$ lattice) is equivalent to a non-commutative 
$U(1)$ theory on a $L^D$ lattice, $L^D\equ N$, is the simplest 
example of the exact Morita equivalence for lattice gauge theories
\cite{amns}.  Note that the number of degrees of 
freedom is the same: $D \; N^2$
for $U(N)$ on the hypercube and $D \; L^D$ for $\cU$ for the 
non-commutative $U(1)$ theory on the $L^D$ lattice.
It illustrates that lattice gauge theory provides us with a 
natural framework for defining 
non-commutative field theories in a {\it non-perturbative} way.

\section{Quantum gravity on the lattice}\label{gravity}

\subsection{General considerations}

As mentioned in the Introduction 
the lattice formalism of dynamical triangulations
seems the ideal framework to address higher dimensional quantum 
gravity. Until now we have not been successful\footnote{see 
however the contribution by Shinichi Horata at this conference for 
interesting progress.}. However, it might 
be that the problems encountered so far are more related to 
{\it Euclidean} gravity than to {\it quantum} gravity. 

In two-dimensional quantum gravity the simplest direct 
approach to the theory of fluctuating geometries has 
been very successful. By rotating to Euclidean signature
and regulating the sum over geometries by introducing 
the reparameterization-invariant lattice cut-off  
called {dynamical triangulations}, one was 
able to calculate generalized Hartle-Hawking wave functionals
and correlation functions depending 
on the geodesic distance \cite{jan2d}.
However, simple generalizations to higher dimensions seem
not to work. While four-dimensional 
quantum gravity may not exist without being embedded in a larger theory,
this is not true for three-dimensional quantum gravity.
This led  to the suggestion \cite{al},
following an old idea by Teitelboim, that Euclidean quantum 
gravity might not be related to ``real'' Lorentzian 
quantum gravity, and that one should only include 
causal geometries in the sum over histories. The geometries which 
appear in the regularized 
version of such a theory were called {\it Lorentzian dynamical 
triangulations}, and each of these geometries has a well defined 
rotation to an Euclidean geometry. The opposite is not true:
there are many Euclidean geometries which cannot be 
rotated to a Lorentzian geometry with a global causal structure.
However, it implies that one {\it can} perform the summation
over this restricted class of geometries in the ``Euclidean sector'',
and rotate back after the summation has been done. This is 
the way we will treat the summation over histories in the following. 
 
In two dimensions one can perform the summation over the 
class of Lorentzian geometries explicitly and  obtain 
a theory which  {\it differs} from Euclidean
two-dimensional quantum gravity. The difference is 
best illustrated by considering what is called the 
proper-time propagator, where one sums over all geometries with 
the topology $S^1\!\times\![0,1]$, where 
the two spatial boundaries are 
separated by a proper time $T$. 
In the Lorentzian theory the spatial  slices at a time $T' < T$ 
are characterized by the fact that the  spatial topology 
always remains a circle. In two-dimensional Euclidean quantum gravity 
similar spatial slices corresponding to constant 
proper time  split up into many (in the continuum limit
infinitely many) disconnected ``baby'' universes, {\it each} having 
the topology $S^1$ \cite{kawai2d}. 

In addition three-dimensional quantum gravity is interesting 
for the following reason: locally, the classical solution 
is just flat space, or in the case of a positive cosmological 
constant, 3d de Sitter space. If one expands around such 
a classical solution in order to quantize the theory
one finds it is non-renormalizable. Nevertheless we know the theory 
has no dynamical {\it field degrees of freedom}, but only
a finite number of degrees of freedom. Thus it can 
be quantized following different procedures, e.g.\ reduced 
phase space quantization. However, it remains unclear if 
anything is ``wrong'' in an approach where one performs the 
summation over fluctuating three-geometries and how 
such an approach deals with the seeming ``non-renormalizability''
of the theory of fluctuating geometries. 

Like in two dimensions, also in three dimensions there will  
be a drastic difference between what we call Euclidean quantum gravity 
and Lorentzian quantum gravity. Loosely speaking, because of the 
restricted class of geometries which enters into the sum
over histories in the case of Lorentzian quantum gravity,
the quantum theory will be better behaved, and,
 contrary to the situation
in a regularized Euclidean quantum gravity theory,
one can define a continuum limit. Seemingly,
Euclidean quantum gravity theory, as defined by 
dynamical triangulations, does not treat the 
conformal factor correctly (see \cite{dl} for a discussion.). It is a little 
surprising since the main success of the formalism in 
two dimensions precisely was the correct treatment of the
conformal mode. The explanation may be related to the
fact that the conformal mode in higher dimensions is 
also the cause of the {\it unboundedness} of the 
Euclidean Einstein action. In the Lorentzian theory 
one still has a conformal mode, but the geometries 
associated with this mode are less ``singular'' than
the geometries one meets in the Euclidean theory 
(see \cite{ajl} for details).

\subsection{3d Lorentzian dynamical triangulations}  

As mentioned above the proper-time propagator is a 
convenient object to study. We choose the simplest possible 
topology of space-time, $S^2\!\times\! [0,1]$, so that the spatial 
slices of constant proper time have the topology of a two-sphere.
Each spatial slice has an induced two-dimensional Euclidean
geometry. In the formalism of Lorentzian dynamical triangulations
the space of Euclidean 2d geometries is approximated by 
the space of 2d dynamical triangulations. This approximation 
is known to work well and in the limit where the lattice spacing 
(the length of the lattice links) goes to zero the continuum 
limit, i.e.\ quantum Liouville theory, is recovered.
In order to obtain a three-dimensional triangulation of 
space-time we have to fill in the space-time between two successive
spatial slices. This is done as follows: above (and below) each 
triangle at proper time $t\equ n\, a$, $n$ an integer,
we erect a tetrahedron with its 
tip at $t\pl a$, a so-called (3,1)-tetrahedron (if the tip is
at $t\mi a$ a (1,3)-tetrahedron). Two tetrahedra which share a 
spatial link in the constant-$t$ plane might be glued together along 
a common time-like triangle. Remaining free time-like triangles
with either the  spatial link  in the constant-time 
slice at $t$ and the tip at 
$t\pl a$, or the spatial link at the constant-time slice at 
$t\pl a$ and the tip at $t$, are 
glued together by so-called (2,2)-tetrahedra. They  have 
a spatial link both in the constant-$t$ slice and the constant 
$t\pl a$ slice.
(2,2)-tetrahedra can also be glued to each other in 
all possible ways, the only restriction being that 
if we cut the triangulation in a spatial plane 
between $t$ and $t\pl a$, the corresponding graph, which 
consists of triangles and squares (coming from cutting the 
(2,2)-simplexes) form a graph with spherical topology.

Summing over all such piecewise linear geometries with the 
Boltzmann weight given by the Einstein-Hilbert action defines
the sum over geometries (see \cite{ajl} for details).
The partition function becomes (up to boundary terms)
\beq\label{wick2}
Z(k_0,k_3,T)=\sum_{T} 
\e^{k_0 N_0(T)-k_3N_3(T)},
\eeq
where the summation is over the class of triangulations mentioned,
$N_0(T)$ denotes the total number of vertices and $N_3(T)$
the total number of tetrahedra in the triangulation $T$. 
$k_0$ is inversely proportional 
to the bare inverse gravitational coupling constant, while 
$k_3$ is linearly related to the cosmological coupling constant.

\subsection{Numerical simulations}

\begin{figure}[htb]
\vspace{-2.5cm}
\includegraphics[width=7.5cm]{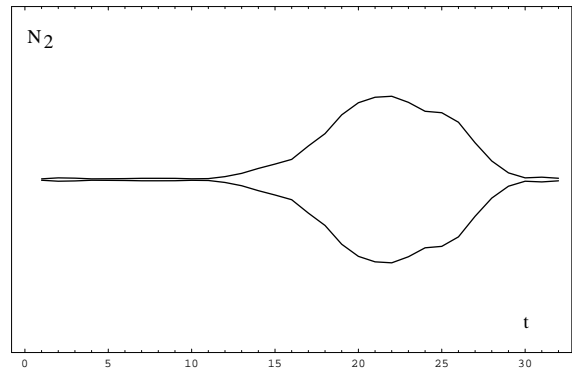}
\vspace{-3.5cm}
\caption[fig3]{A snapshot of a three-dimensional configuration of 
space-time. The vertical axis is the spatial volume $N_2$, the 
horizontal axis the  proper time $t$.}
\label{fig3}
\end{figure}

The model \rf{wick2} can be studied by Monte Carlo simulations
(see \cite{ajl} for details). There is only one phase\footnote{In
some previous studies we observed a phase transition for large 
$k_0$. This was caused by restrictions on the gluing of 
(2,2)-simplices. We have now dropped these restrictions.}.
Let us fix the total three-volume of space-time to be $N_3$,
and let us take the total proper time $T$ large. 
One observes the appearance of an ``semiclassical'' lump of universe,
as shown in Fig.\ \ref{fig3}.
As we change  the bare gravitational coupling constant,
the time-extent of the semiclassical lump decreases. However, 
all correlation functions we have studied can be mapped into 
each other by a simple rescaling of time and space directions. 
Further, we observe that a typical  spatial volume $N_2(t)$ in 
the lump and the time-extent $\Del T$ of the lump scale as:
\beq\label{scale}
N_2 \sim N_3^{2/3},~~~~\Del T \sim N_3^{1/3}.
\eeq
This justifies the use of the word ``semiclassical'' in the 
description of the lump.
In the computer simulations we observe that 
the center of mass of the lump 
moves around randomly. In addition there are  fluctuations in the 
spatial volume. We have studied the fluctuations of 
{\it successive} spatial volume. The  distribution 
of such spatial volumes is very well described by the formula
\beq\label{distrib}
P(N_2(t),N_s(t+a)) \sim 
\e^{ -c(k_0) \frac{(N_2(t\pl a)-N_2(t))^2}{N_2(t\pl a)+N_2(t)}}.
\eeq
The constant $c(k_0)$ decreases as $k_0$ increases (i.e.\ the 
bare gravitational coupling constant decreases). At the same 
time one can observe that the total number of (2,2)-simplices
decreases, indicating that the (2,2)-simplices act as glue between successive 
spatial volumes.

Thus the leading terms in the  
effective action for the spatial volume of the model
are given by 
\beq\label{sitter} 
S_{eff}(N_2) = \int dt \; \Big( \frac{\dot{N}_2^2(t)}{N_2(t)} + \L N_2(t)\Big).
\eeq
This is exactly the classical Lorentzian action for the spatial 
volume in proper-time gauge, thereby supporting the 
semiclassical interpretation of the lump.

\subsection{3d Lorentzian gravity as a matrix model}

\begin{figure}[htb]
\vspace{-1.5cm}
\centerline{\includegraphics[width=8.0cm,angle=-90]{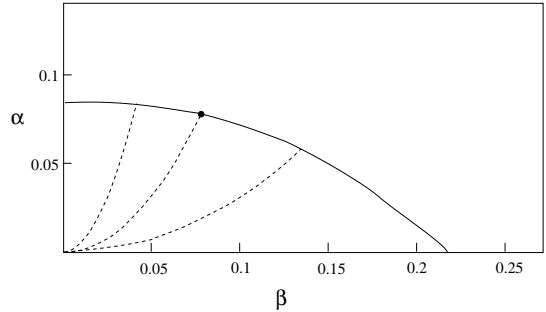}}
\vspace{-2.5cm}
\caption[fig4]{The phase diagram of 3d Lorentzian quantum 
gravity from matrix models}
\label{fig4}
\end{figure}

If we slice our three-dimensional configurations, not at proper
time $t$, but at proper time $t\pl a/2$ we will, as mentioned 
earlier, obtain a spherical graph with two types of triangles,
coming from the spatial intersections 
of (1,3)- and (3,1)-tetrahedra, respectively.
In addition the graph will contain 
squares coming from the (2,2)-tetrahedra. This class of graphs 
can be described by a two-matrix model:
\beq\label{4.1}
Z \equ \int \!\!\d A \d B \, \e^{ \mi N \tr \left( (A^2\pl B^2) 
\mi \a (A^3 \pl B^3)\mi \b ABAB\right)}.
\eeq
$A$ and $B$ are $N\!\times\! N$ Hermitian matrices and the spherical
graphs are selected in the large-$N$ limit. The coupling 
constants $\a,\b$ can be related to the gravitational coupling 
constants. While this matrix model has not been solved, 
there exists another, closely related matrix model where
\beq\label{4.2}
A^3+B^3 \to A^4+B^4,
\eeq
which {\it has} been solved \cite{kz}. This 
model has a simple interpretation in terms of ``triangulations'':
the (3,1)- and (1,3)-tetrahedra are replaced by (4,1)- and (1,4)-pyramids
in an obvious notation. This model, where the building blocks are 
pyramids and (2,2)-tetrahedra, is an equally good regularization 
of three-dimensional gravity. Again one can work out the 
relation between the coupling constants $\a,\b$ and $k_0,k_3$
(see \cite{ajlv} for details). The matrix model is 
defined for sufficiently small values of $\a,\b$ and  has 
a critical line in the $(\b,\a)$-plane where the continuum 
limit is obtained, see Fig.\ \ref{fig4}. 
The three dotted curves approaching the critical line 
correspond to three different fixed 
values of the bare gravitational constant. The 
bare cosmological constant $k_3$ changes along the dotted lines 
to achieve its critical value $k_3^c(k_0)$ 
(which depend on $k_0$) at the 
critical line shown in the figure. Fine tuning of $k_3$ to
$k_3^c(k_0)$ corresponds to a renormalization of the cosmological 
constant and the approach to the infinite volume limit 
of the theory: for $k_3 \to k_3^c(k_0)$ the expectation value 
of the number of tetrahedra or pyramids diverges. On the other 
hand there seems to be no need for a renormalization of $k_0$:
it simply defines an overall scale for the model.
 
If we follow the critical line of the 
matrix model, starting at small values of $\b$, it corresponds initially  
to a weak-coupling phase of three-dimensional quantum gravity where 
the bare gravitational coupling constant is small ($k_0$ large).
This is the phase we have observed in the computer simulation.
However, the matrix model undergoes a phase transition at  $\a \equ \b$,
shown by a dot on Fig.\ \ref{fig4}, separating the weak-coupling 
phase from a strong-coupling phase, corresponding to large values 
of the gravitational coupling constant. In this phase the 
triangulations corresponding to the spatial slices at $t$ and $t\pl a$
of topology $S^2$ disintegrate into many (in the continuum limit
 where $a \!\to\! 0$ into infinitely many) baby universes
each of topology $S^2$, connected by a web of thin wormholes. 
This is possible because the matrix 
model admits more general configurations than were allowed in 
the computer simulations. The only requirement of the matrix 
model is that the combined graph at $t\pl a/2$ is spherical.
The component coming from the spatial slice at $t$ 
can actually be disconnected (and similar for the component coming from the 
spatial slice at $t\pl a$). 
It becomes a dynamical question what happens if one allows
for such fluctuations.
In the weakly coupled phase the spatial topology stays 
unchanged as $S^2$, but increasing the gravitational coupling
constant space starts to be  torn apart. When the gravitational coupling 
constant is sufficiently large a phase transition takes place and 
space disintegrates into 
baby universes, only connected by thin wormholes
(see \cite{ajlv} for details). In this 
way the model provides a  concrete realization of the 
ideas of Wheeler  and Hawking of a quantum foam at short distances.

Defined in a non-perturbative
way on the lattice, three-dimensional quantum gravity reveals a 
rich structure, and, as in two dimensions, the model can 
be analyzed by a fruitful interplay between numerical and analytic methods.

Hopefully, the same will be true in the case of four-dimensional 
Lorentzian quantum gravity defined via dynamical triangulations.

\end{document}